\documentclass[twocolumn,superscriptaddress,longbibliography,aps,
pra,preprintnumbers,10pt]{revtex4-2}

\usepackage[normalem]{ulem}
\usepackage{graphicx}
\usepackage{bm}
\usepackage{color}
\usepackage{epstopdf}
\usepackage{amsmath}
\usepackage{amssymb}
\usepackage{epstopdf}
\usepackage{multirow}
\usepackage[urlcolor=blue,colorlinks=true,citecolor=blue,linkcolor=blue,pdfstartview={FitH},bookmarks=false]{hyperref}
\usepackage{xcolor}

\begin{document}

\title{Connection between the semiconductor--superconductor transition\\ and the spin-polarized superconducting phase in the honeycomb lattice}

\author{Agnieszka Cichy}
\email[e-mail: ]{agnieszkakujawa2311@gmail.pl}
%\homepage[ORCID ID: ]{https://orcid.org/0000-0001-5835-9807}
\affiliation{\mbox{Institute of Spintronics and Quantum Information, Faculty of Physics}, 
Adam Mickiewicz University in Pozna\'n, 
Uniwersytetu Pozna\'{n}skiego 2, 61-614 Pozna\'{n}, Poland}
\affiliation{Institut f\"{u}r Physik, Johannes Gutenberg-Universit\"{a}t Mainz, Staudingerweg 9, 55099 Mainz, Germany}

\author{Konrad Jerzy Kapcia}
\email[e-mail: ]{konrad.kapcia@amu.edu.pl}
%\homepage[ORCID ID: ]{https://orcid.org/0000-0001-8842-1886}
\affiliation{\mbox{Institute of Spintronics and Quantum Information, Faculty of Physics}, Adam Mickiewicz University in Pozna\'n, 
Uniwersytetu Pozna\'{n}skiego 2, 61-614 Pozna\'{n}, Poland}
\affiliation{\mbox{Center for Free-Electron Laser Science CFEL, Deutsches Elektronen-Synchrotron DESY}, Notkestr. 85, 22607, Hamburg, Germany}

\author{Andrzej Ptok}
\email[e-mail: ]{aptok@mmj.pl}
%\homepage[ORCID ID: ]{https://orcid.org/0000-0002-5566-2656}
\affiliation{\mbox{Institute of Nuclear Physics, Polish Academy of Sciences}, 
W. E. Radzikowskiego 152, 31-342 Krak\'{o}w, Poland}

\date{\today}

\begin{abstract}
The band structure of noninteracting fermions in the honeycomb lattice exhibits the Dirac cones at the corners of the Brillouin zone.
As a consequence, fermions  in  this lattice manifest a se\-mi\-con\-duc\-ting behavior below some critical value of the onsite attraction, $U_{c}$.
However, above $U_{c}$, the superconducting phase can occur.
We discuss an interplay between the semiconductor--superconductor transition and the possibility of realization of the spin-polarized superconductivity (the so-called Sarma phase).
We show that the critical interaction can be tuned by the next-nearest-neighbor (NNN) hopping in the absence of the magnetic field.
Moreover, a critical value of the NNN hopping exists, defining a range of parameters for which the semiconducting phase can emerge.
In the weak coupling limit case, this quantum phase transition occurs for the absolute value of the NNN hopping equal to one third of the hopping between the nearest neighbors.
Similarly, in the presence of the magnetic field, the Sarma phase can appear, but only in a range of parameters for which initially the semiconducting state is observed.
Both of these aspects are attributed to the Lifshitz transition, which is induced by the NNN hopping as well as the external magnetic field.
\end{abstract}

\maketitle

\section{Introduction}

The realization of the honeycomb lattice in graphene draws a lot of attention of the scientific community~\cite{beenakker.08,castroneto.guinea.09,dassarma.adam.11,goerbig.11,basov.fogler.14,avsar.ochoa.20}.
The extraordinary properties of the honeycomb la\-tti\-ce are mainly associated with massless Dirac fermions, which are located in the corners of the Brillouin zone~\cite{novoselov.geim.05}.
This lattice exhibits also topological properties manifested by the existence of zero-energy edge states~\cite{fujita.wakabayashi.96,nakada.fujita.96,wakabayashi.sigrist.00} or in the quantum Hall effect~\cite{zhang.tan.05,novoselov.mccann.06,novoselov.jiang.07}, associated to the finite Berry curvature in these systems~\cite{xiao.chang.10}.
The electronic properties of the honeycomb lattices (in graphene or in related two-dimensional materials) has opened new avenues of research in which applications play a very important role, i.e., spintronics~\cite{avsar.ochoa.20} or valleytronics~\cite{bussolotti.kawai.18}.

The relatively simple way of manipulating real ho\-neycomb layers (e.g., in graphene or transition metal dichalcogenides) enabled progress in realization of the Moir\'{e} twisted bilayer lattices~\cite{bistritzer.macdonald.11}.
The most in\-te\-res\-ting phenomena observed in these systems are, among other things, unconventional superconductivity~\cite{cao.fatemi.18,lu.setpanov.19,yankowitz.shaowen.19,arora.polski.20,codecido.wang.19,saito.ge.20,stepanov.das.20,saito.ge.20} as well as  an  insulating phase~\cite{cao.fatemi.18b,saito.ge.20,stepanov.das.20,codecido.wang.19}, the topological edge states~\cite{zhang.macdonald.13,vaezi.liang.13,ju.shi.15,brown.walet.18}, or fractional quantum Hall effect~\cite{hunt.sanchezyamagishi.13}.

\paragraph*{Motivation.}
The presence of the Dirac points in the honeycomb lattice band structure has important consequences related to the physical properties of the systems.
One of them is the realization of the se\-mi\-con\-duc\-ting phase at half-filling, below some critical interaction $|U_{c}|/t \sim 2.1 \div 2.3$~\cite{zhao.paramekanti.06,cichy.ptok.18,iskin.19}.
In this case, 
the se\-mi\-con\-duc\-ting behavior is related to: (i) vanishing density of states at the Fermi level and (ii) two bands (the conduction and valence bands), which touch each other at some points in momentum space (i.e., the Fermi surface shrinks to the Dirac points). However,
above the critical attraction, the superconducting phase is stable.
Hence, one can observe a semiconductor--superconductor transition, which does not take place away from half-filling~\cite{zhao.paramekanti.06}.
However, including the next-nearest-neighbor (NNN) hopping leads to a changed value of $U_{c}$~\cite{iskin.19}.

\begin{figure*}
\includegraphics[width=\linewidth]{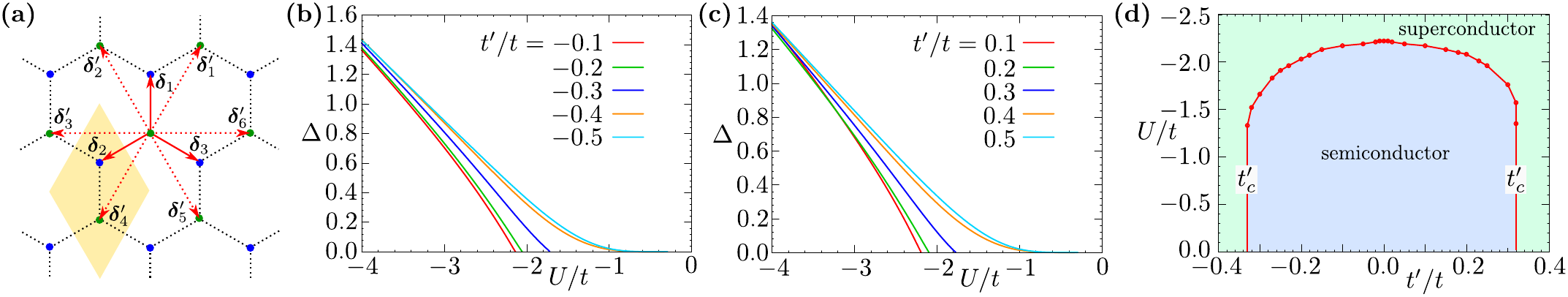}
\caption{%
(a) Schematic representation of the honeycomb lattice.
The positions of nearest- and  next-nearest sites are indicated by solid red and dashed red arrows, respectively.
The role of the onsite attraction on the stabilization of the superconducting phase at half-filling for (b) negative and (c) positive $t'$.
(d) The semiconductor--superconductor phase diagram presenting the dependence of the critical interaction $U_{c}$ (solid red line) on the next-nearest-neighbor hopping ($t'$) and revealing the existence of the critical values $| t'_{c} | = t/3$. 
For $|t'|>|t'_c|$ the semiconducting behavior disappears.
The results are obtained in the absence of the magnetic field at half-filling ($n=1$) and $T = 0$.
\label{fig.1}
}
\end{figure*}

Here, we study the predominant role of the NNN hopping in the semiconductor--superconductor phase transition.
These studies give important knowledge about physics of systems in which the Dirac fermions are realized.
Relevant examples of such systems are superconductors containing the honeycomb sublattices (e.g., FePSe$_{3}$~\cite{wang.ying.18}, CrPS$_{4}$~\cite{susilo.jang.20}, CrSiTe$_{3}$~\cite{cai.sun.20}, SnPS$_{3}$~\cite{yue.zhong.21}, or Cu$_{2}$I$_{2}$Se$_{6}$~\cite{cai.lin.19}) as well as certain transition metal dichalcogenides~\cite{saito.nakamura.16}.
However, also interfaces between Dirac semimetals and superconductors can reveal similar properties~\cite{huang.zhou.19,zhu.wang.20,grabacki.dabrowski.20,kononov.edres.21,lee.inturu.21}.
Moreover, recent progress in the experimental realization of artificial lattices, e.g., the artificial honeycomb lattice created within optical lattices~\cite{tarruell.greif.12,uehlinger.jotzu.13,mei.zhang.13} or by atomic nanostructures~\cite{gomes.mar.12}, leads also to great opportunities of studying the described unique phenomena.
Finally, we discuss also the influence of the semiconductor--superconductor transition on possible stabilization of the spin-polarized superconducting phase ({\it Sarma phase})~\cite{sarma.63}.
We conclude that both properties can be explained by the Lifshitz transition induced by the NNN hopping or the magnetic field.

\paragraph*{Theoretical background.}
In this work, we investigate an {\it s}-wave superconductivity on a honeycomb lattice with NNN hopping [Fig.~\ref{fig.1}(a)].
The system is described by the following Hamiltonian: $H = H_\text{kin} + H_\text{sc}$, where
\begin{eqnarray}
H_\text{kin} = \sum_{ij\sigma} \left[-t_{ij} - \left( \mu + \sigma h \right) \delta_{ij} \right] c_{i\sigma}^{\dagger} c_{j\sigma} ,
\end{eqnarray}
describes the kinetic part (cf. also Appendix~\ref{sec.wypr}).
Here, $c_{i\sigma}$ ($c_{i\sigma}^{\dagger}$) denotes the a\-nni\-hi\-la\-tion (creation) operator of a fermion with spin $\sigma$ in the
\mbox{$i$-th} site, $\mu$ is the chemical potential, while $h$ is the external magnetic field.
In our consideration, we assume that the particles can hop between nearest neighbors (with hopping integral $t_{ij} \equiv t > 0$
as energy unit) and next-nearest neighbors (with hopping integral $t_{ij} \equiv t'$ as free parameter).

In turn, the source of the superconducting phase is the pairing interaction in the form of the Coulomb term $U \sum_{i} n_{i\uparrow} n_{i\downarrow}$ (where $n_{i\sigma}=c_{i\sigma}^{\dagger} c_{i\sigma}$ is the particle number operator), which after the mean-field decoupling leads to the BCS-like term:
\begin{eqnarray}
H_\text{sc} = U \sum_{i} \left( \Delta_i c_{i\downarrow}^{\dagger} c_{i\uparrow}^{\dagger} + H.c. \right) - U \sum_{i} | \Delta_{i} |^{2} ,
\end{eqnarray}
where $U < 0$ describes the on-site attraction  between particles with opposite spins on the same site, while $\Delta_{i} = \langle c_{i\downarrow} c_{i\uparrow} \rangle$ is the superconducting order parameter (here, we assume $\Delta_i \equiv \Delta$ due to the consideration of a spatially homogeneous system).
The ground state can be found from minimization of the grand canonical potential $\Omega \equiv \Omega (\Delta) = - k_{B} T \text{ln} \{ \text{Tr} [ \exp ( - H / k_{B} T ) ] \} $ with respect to $\Delta$, for fixed parameters $\mu$, $t'$, and $U$.
This allows us to find the superconducting order parameter (from the gap equation \mbox{$\partial \Omega / \partial \Delta = 0$}), the total number of particles (\mbox{$n = - \partial\Omega/\partial\mu$}), and the magnetization (\mbox{$m = - \partial\Omega/\partial h$}).
All details of the analytical derivation can be found in Ref.~\cite{cichy.ptok.18}.

\section{Results}

\paragraph*{Semiconductor--superconductor transition.}
As mentioned above, the increase of the pairing interaction $U$ leads to a phase transition from the semiconducting to the superconducting phase~\cite{zhao.paramekanti.06,cichy.ptok.18,iskin.19}.
This transition is manifested by a change of the $\Delta = \Delta(U)$ functional form [cf. Fig.~\ref{fig.1}(b) and Fig.~\ref{fig.1}(c)].
When the semiconducting phase can be realized, $\Delta$ drops to zero at some finite value of $U=U_c$ (for $|t'|<|t'_{c}|$)~\cite{kopnin.sonin.08}.
Otherwise, $\Delta$ decays exponentially to zero with decreasing $|U|$ (for $|t'|>|t'_{c}|$)~\cite{ptok.crivelli.15}.
This change in the $\Delta(U)$ behavior leads to the phase diagram presented in Fig.~\ref{fig.1}(d).
For $|t'| < |t'_{c}|$, the semiconductor--superconductor phase transition exists and $U_{c}$ (marked by solid red line) has a~finite value.
However, for $|t'|$ above some critical value of the NNN hopping [i.e., $t'_{c}$, marked by the dashed line in Fig.~\ref{fig.1}(d)], this transition does not occur and only the superconducting phase exists (for $|t'| > |t'_{c}|$, $U_{c}=0$).
The $U$-dependence of the order parameter $\Delta$ gives information about the type of semiconductor--superconductor phase transition [see Fig.~\ref{fig.1}(b) and Fig.~\ref{fig.1}(c)].
This transition is of the second order because $\Delta$ changes continuously at the transition boundary, although there could be a discontinuous change of $U_c$ with the change of $t'$.
Described properties show the crucial influence of $t'$ parameter on the semiconductor--superconductor transition.
Indeed, in the next paragraph, we will show that the existence of $t'_{c}$ has important impact on physical properties of the system, both with and without the external magnetic field.

\begin{figure*}
\includegraphics[width=\linewidth]{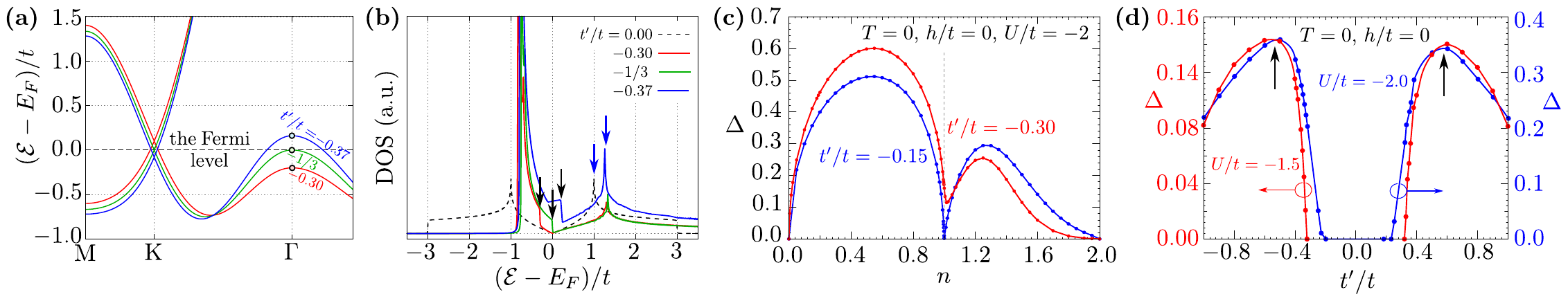}
\caption{
Band structure along high-symmetry-point paths in the Brillouin zone (a) and density of states (b) of the non-interacting system presenting the emergence of the Lifshitz transition near the critical value of the hopping between the next-nearest neighbors $t'$ (as labeled).
Black arrows show jumps of DOS indicating the Lifshitz transition induced by $t'$, while blue arrows show positions of the van Hove singularities.
The dependence of the superconducting gap on the filling $n$ (c) and on the next-nearest-neighbor hopping  $t'$ (d) for fixed parameters (as labeled, details in text).
\label{fig.2}
}
\end{figure*}

\paragraph*{Role of the Lifshitz transition.}
The occurrence of the transition from the semiconducting to the superconducting phase (for fixed $U$) is strongly associated with the band dispersion at the high symmetry points of the Brillouin zone.
The non-zero hopping $t'$ leads to the mo\-di\-fi\-ca\-tion of the dispersion relation of the bands, presented in Fig.~\ref{fig.2}(a).
For $| t' | = | t'_{c} |$, the Lifshitz transition takes place~\cite{lifshitz.60} and the Dirac point (located at K point), does not separate the valence band (fully occupied) from the conduction band (fully empty) at half-filling ($n=1$).
In general, the Lifshitz transition is characterized by changes of the Fermi surface (FS) topology due to the variation of the Fermi energy or/and the band structure.
In this particular case, the Lifshitz transition is associated with the crossing of the Fermi level by the band at $\Gamma$ point [marked by dots in Fig.~\ref{fig.2}(a)] and emergence of the new FS.
Hence, the exact value of $t'_{c}$ can be found analytically as $\pm t/3$ (for all the details of the analytical derivation see: Appendix~\ref{sec.wypr}).
This value of $| t'_{c} |$ does not change for the weak coupling limit [see Fig.~\ref{fig.1}(d)].

The Lifshitz transition is also reflected in the density of states (DOS) properties in the normal state [Fig.~\ref{fig.2}(b)].
For simplification of the DOSs comparison, we define the Fermi level ($E_{F}$) related to the half-filing to be at zero energy.
The DOS of the honeycomb lattice at $t'=0$ (dashed black line) exhibits a symmetric form with respect to the center of the bandwidth (i.e., the Dirac points).
The increase of $| t' | \ll |t'_c|$ causes that DOS is asymmetric, without destroying the Dirac points.
However, for $| t' |  = | t'_c |$, the Lishitz transition leads to a non-zero value of the DOS at the Fermi level for half-filling.
Black arrows in Fig.~\ref{fig.2}(b) denote the jump of DOS associated to the local maximum at the $\Gamma$ point [marked by circles in Fig.~\ref{fig.2}(a)].
At the same time, the position of the van Hove singularity [cf. blue arrows in Fig.~\ref{fig.2}(b)] is only slightly affected.

In the weak coupling limit, $|t'_c|$ does not depend on the pairing interaction and is equal to $t/3$ [cf. Fig.~\ref{fig.1}(d)].
With increasing $|U|$, the suppression of the se\-mi\-con\-duc\-ting phase is observed, and there are deviations from $t'_c$ calculated analytically for the noninteracting system.
For instance, at fixed pairing interaction \mbox{$|U|/ t = 2$} and $| t' | \ll |t'_c|$, two superconducting domes can be distinguished [blue line in Fig.~\ref{fig.2}(c)].
The dependence of the superconducting order parameter on the filling is reflected in the DOS asymmetry for $t' \neq 0$.
Here, the semiconducting behavior at half-filling is manifested by the vanishing of $\Delta$.
For $t'$ around $t'_{c}$, the superconducting phase exists in the whole range of particle concentration. Finally, for stronger couplings, i.e., for $| U | \geq |U_{c}|$, the semiconducting phase is unstable.
The suppression of the semiconducting state is associated with the BCS-BEC crossover~\cite{zhao.paramekanti.06}.

\begin{figure}[!t]
\includegraphics[width=\linewidth]{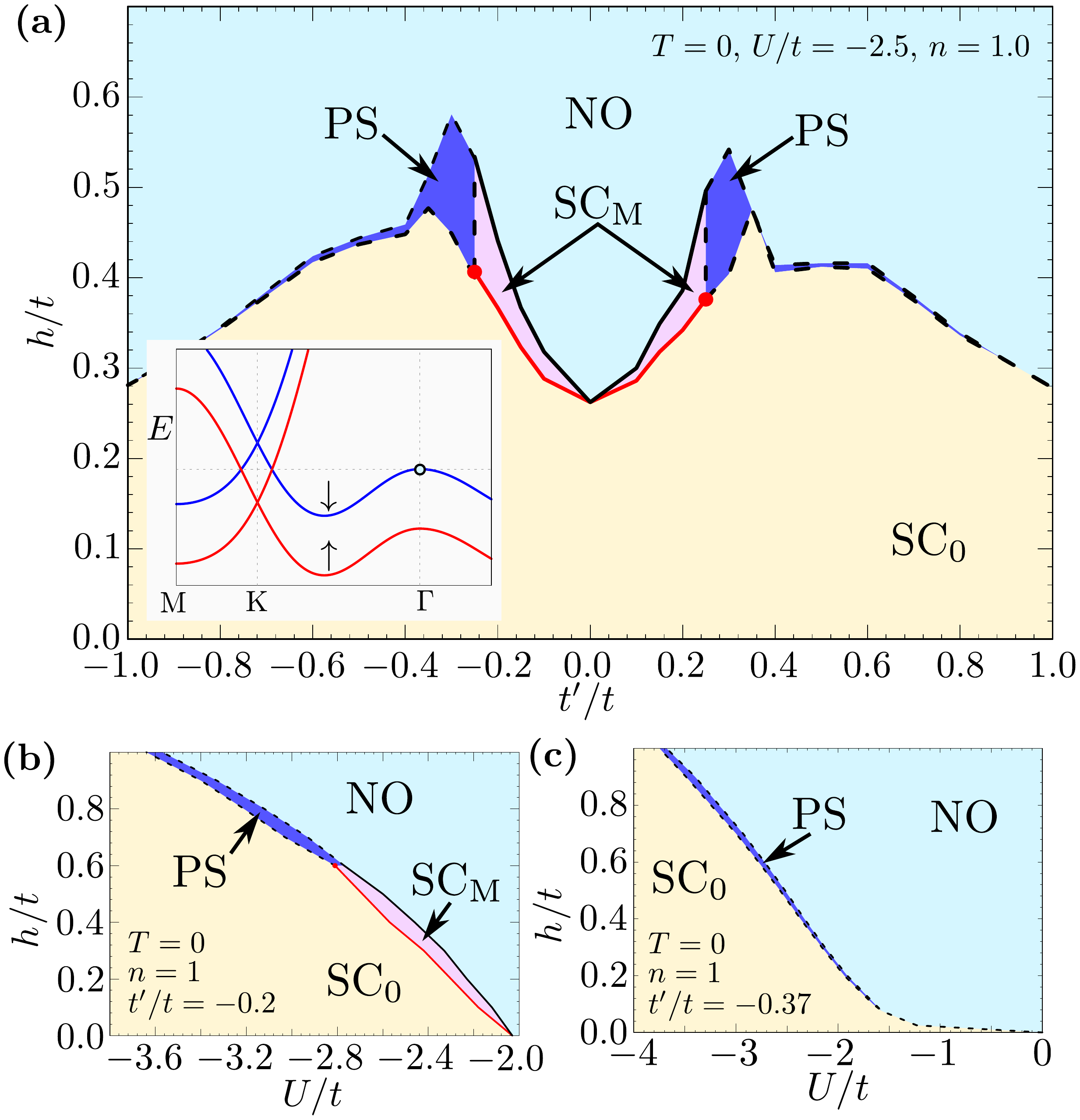}
\caption{
(a) The $h$--$t'$ phase diagram presenting the transition from the superconducting to the normal state for $| U | /t = 2.5$ at half-filling.
The phases are labeled as: NO -- normal state, SC$_{0}$ -- BCS-like phase, SC$_\text{M}$ -- spin-polarized superconducting phase, and PS -- phase separation.
Panels (b) and (c) present $h$--$U$ phase diagrams for fixed $t'$.
The inset in the panel (a) shows the band structure of the non-interacting system in the presence of the external magnetic field $h \neq 0$.
Red and blue lines correspond to the spin $\uparrow$ and $\downarrow$ of the electrons, respectively.
\label{fig.3}
}
\end{figure}

Typically, the Lifshitz transition leads to the mo\-di\-fi\-ca\-tion of the critical parameters of the superconducting phase~\cite{shi.han.17,ptok.rodriguez.18}.
This is related to the occurrence of additional peaks in the DOS.
For relatively small $|U|$, when transition from the semiconducting to the su\-per\-con\-duc\-ting phase takes place at $| t' | = |t_{c}|$, no significant change of the superconducting gap is observed [Fig.~\ref{fig.2}(d)].
However, when the superconducting phase exists for $t'$ around the critical value $t'_{c}$ (e.g., \mbox{$|U|/t = 2.0$} indicated by the blue solid line), an increase of the superconducting order parameter is observed.
The emergence of the superconducting phase for $| t' | = | t'_{c} |$ is allowed by the non-zero DOS at the Fermi level.
At the same time, the maximum value of the superconducting order parameter $\Delta$ occurs for $| t' | > |t'_{c}|$ [marked by black arrows in Fig.~\ref{fig.2}(d)].
Similar behavior of the critical magnetic field $h_c$ in some range of $t'$ around $t'_c$ is also observed [Fig.~\ref{fig.3}(a)].
In this case, the critical magnetic field exhibits the maximum value for $t'$ close to $t'_{c}$.

\paragraph*{Realization of the spin-polarized superconducting phase.}
The transition from the semiconducting to the superconducting phase can be also driven by the pairing interaction $U$.
For instance, as we mentioned above, in the absence of the NNN hopping ($t'=0$), the semiconducting phase is stable below the critical interaction $U_{c}$ [cf. Fig.~\ref{fig.1}(d)].
However, even for the pairing interaction above $U_{c}$, where the superconducting phase is stable independently from $t'$, the critical value of the NNN hopping $t'_{c}$ plays an important role in the physical properties of the system (e.g., in the presence of the external magnetic field).

In general, the magnetic field can act on the system in two different ways: by the orbital and paramagnetic effect.
The orbital (diamagnetic) effect leads to the occurrence of the Abrikosov vortex state~\cite{abrikosov.57}.
The paramagnetic (Pauli) pair-breaking effect originates from the Zeeman splitting of the electron energy levels.
The relation between these two, above mentioned contributions is described by the Maki parameter $\alpha = \sqrt{2} H_\text{c2}^\text{orb} / H_\text{c2}^{P}$~\cite{maki.66}, where $H_\text{c2}^\text{orb}$ and $H_\text{c2}^{P}$ are the critical magnetic fields at $T=0$, derived from the orbital and diamagnetic effects, respectively.
Typically, the diamagnetic effect is responsible for destroying the superconducting state (type-II superconductors), which corresponds to $\alpha < \sqrt{2}$.
However, there exists some class of matrials for which the diamagnetic effect is negligible and the paramagnetic effect plays the crucial role (i.e., $\alpha > \sqrt{2}$).
These  systems are called the Pauli systems and exhibit the Chandrasekhar--Clogston limit~\cite{chandrasekhar.62,clogston.62}.
Such a limit is realized, e.g., in the heavy-fermion systems~\cite{matsuda.shimahara.07}, organic superconductors~\cite{wosnitza.15}, or iron-based superconductors~\cite{zocco.grube.13,cho.yang.17,kasahara.sato.20} (i.e., typically layered compounds with a weak coupling between the layers).
Nevertheless, the role of the orbital effects can be reduced (or even negligible), when the magnetic field is directed in the plane of the layer.
Moreover, an existence of an interface between superconductor and ferromagnetic material~\cite{efetov.garifullin.08,burch.mandrus.18,gong.zhang.19,kezilebieke.huda.20} can lead to the similar effect.
In our study, we are focused only on the paramagnetic effects introduced by the Zeeman magnetic field.

Let us start with a discussion of the influence of the magnetic field on the superconducting phase (Fig.~\ref{fig.3}).
For $| U | = 2.5 t > | U_c |$, the BCS-like superconducting phase (SC$_{0}$) is stable in the whole region of $t'$ [see Fig.~\ref{fig.1}(d)].
On the other hand, in the presence of a magnetic field, the system undergoes a discontinuous phase transition from the superconducting to the normal phase.
However, around this transition, a superconducting spin-polarized phase (SC$_\text{M}$), i.e. the Sarma phase~\cite{sarma.63}, has been found as well as the phase separation (PS) region between the superconducting and the normal state.
The SC$_\text{M}$ state is the spatially homogeneous superconductivity which has a gapless spectrum for the majority spin species~\cite{gubbels.romans.06}.
Moreover, the boundary between SC$_\text{M}$ and PS is restricted by $| t' | = | t'_{c} |$.
In fact, the Sarma phase can be stabilized only in a range of $t'$ for which the semiconducting state is realized for the lower values of attractive interactions.
For $|t'| < |t'_c|$ [Fig.~\ref{fig.3}(b)], the SC$_\text{M}$ exists below some value of $|U|$ around $h_{c}$.
However, for $|t'| > |t'_c|$ [Fig.~\ref{fig.3}(c)], the  phase separation region exists between the NO and the SC$_0$ %(BCS-like) 
phases.
This behavior suggests an important role of $t'$ in the realization of the spin-polarized superconducting phase.

In the context of the above analysis, it is worth mentioning that the Lisfhitz transition can be induced by the external magnetic field~\cite{ptok.cichy.17,ptok.kapcia.17}.
A similar si\-tu\-a\-tion takes place in the system under consideration.
It becomes clear when we analyze the behavior of the band structure in the presence of the magnetic field $h$ [cf. inset in Fig.~\ref{fig.3}(a)].
For relatively low values of $h$, the splitting of the band for particles with opposite spins is observed (red and blue lines, respectively).
The Fermi level (the horizontal line) is located at the energy, 
where the bands for spin $\uparrow$ and $\downarrow$ are crossed (for momenta ${\bm k}$ near the K point).
However, when the magnetic field is large enough, the band at the $\Gamma$ point plays an important role.
Indeed, for some magnetic field, the band of electrons with spin $\downarrow$ starts crossing the Fermi level at the $\Gamma$ point.
Then, the spin imbalance can be realized in the system due to the emergence of the third fully spin-polarized Fermi surface pocket which plays a key role in the occurrence of the Sarma phase in the system.
This fact shows again that not only semimetal-superconductor transition is needed to make SC$_\text{M}$ phase stable but also $t'\neq 0$, which, in the presence of magnetic field, generates the extra FS pocket in the system.

It is worth mentioning that, according to some previous investigations~\cite{parish,parish2,ACichyEPL,ACichyAoP}, at $T = 0$, the Sarma phase is unstable in the weak coupling limit for the {\it s}-wave pairing symmetry, both for the continuum model with contact attraction~\cite{parish,parish2} and for the spin-polarized attractive Hubbard model, for different lattice geometries~\cite{ACichyEPL,ACichyAoP}.
Hence, both the unique properties of the honeycomb lattice and the NNN hopping are crucial to stabilize the Sarma phase.

\section{Summary and outlook}

The honeycomb lattice exhibits unique physical properties because of its band structure, in which two bands touch each other in the Dirac cones.
Hence, if we take only the nearest-neighbor hopping $t$ into account ($t'=0$), there is a quantum phase transition at some critical pai\-ring interaction $U_c$~\cite{zhao.paramekanti.06}.

In this work we have discussed and revealed the influence of the next-nearest-neighbor hopping $t'$ on the semiconductor--superconductor quantum phase transition.
We found that there exists some critical value of $t'$ for which the semiconducting behavior disappears and only the superconducting phase becomes stable.
The existence of the additional critical parameter which describes the semiconductor--superconductor transition is related to the Lifshitz transition, which takes place when $| t' |$ increases.
The Lifshitz transition is associated to the part of the band which crosses the Fermi level at the $\Gamma$ point.
Therefore, the exact value of the critical $t'$ can be found analytically in the case of relatively low values of pairing interaction (weak coupling limit) and its absolute value is $t/3 $.
Additionally, we have found that the spin-polarized superconducting Sarma phase occurs in the presence of an external magnetic field.
We have also  attributed this phenomenon to the Lifshitz transition which is induced by the external magnetic field and it leads to the emergence of an extra fully spin-polarized Fermi surface pocket around the $\Gamma$ point.
The Sarma phase can be stable at half-filling and only for $|t'|$ smaller than the critical value $|t'_c|=t/3$.

The presented results explain basic properties of the systems on the honeycomb lattice, and directly establish conditions necessary for realization of the Sarma phase.
In fact, the stability of the Sarma phase is still under debate.
In this context, experiments with ultracold atoms as quantum simulators~\cite{schafer.fukuhara.20} give great opportunities to confirm our predictions of the Sarma phase realization in the honeycomb lattice with the next-nearest-neighbor hopping.

\appendix

\section{Exact value of $t'_{c}$ for the weak coupling limit}
\label{sec.wypr}

In the \mbox{$U\rightarrow 0$} limit, the value of $t'_c$ can be determined from the band structure analysis of the noninteracting system. In this case, the dispersion relation is found as: \mbox{$\mathcal{E}^{\pm}_{{\bm k}\sigma} = \pm t | g_{\bm k} | - t' f_{\bm k} - ( \mu + \sigma h )$}, where
\mbox{$g_{\bm k} = \sum_{i=1}^{3} \exp \left( i {\bm k} \cdot \delta_{i} \right)$} and 
$f_{\bm k} = \sum_{i=1}^{6} \exp \left( i {\bm k} \cdot \delta'_{i} \right)$,
while $\delta_{i}$ and $\delta'_{i}$ denote the real space vectors describing the position of NN and NNN sites, respectively [see Fig.~\ref{fig.1}(a)].
More precisely, $| g_{\bm k} | = \sqrt{3 + f_{\bm k}}$, and $f_{\bm k} = 2 \cos \left( \sqrt{3} k_x \right) + 4 \cos \left( \sqrt{3}k_x/2 \right) \cos \left( 3k_y/2\right)$.
The sign in the first term corresponds to the upper ($+$) and lower ($-$) band.
Therefore, the energy value at the high symmetry points can be found as: $\mathcal{E}^{\pm}_{\text{K}\sigma} = 3t'-\mu$ and $\mathcal{E}^{\pm}_{\Gamma\sigma} = \pm 3 t - 6 t'-\mu$.
The condition for the Lifshitz transtion is \mbox{$\mathcal{E}^{\pm}_{\text{K}\sigma} = \mathcal{E}^{\pm}_{\Gamma\sigma}$}, which gives directly the critical value of \mbox{$| t'_{c} | = t/3$}.
This value of $| t'_{c} |$ does not change for the weak coupling limit.

\vspace*{0.5cm}

\begin{acknowledgments}
We kindly thank the late Roman Micnas for all his inspiration and motivation that he provided for this work. 
We also thank Krzysztof Cichy for careful reading of the manuscript, valuable comments, and discussions.
This work was supported by National Science Centre (NCN, Poland) under Project
Nos. UMO-2017/24/C/ST3/00357 (A.C.), UMO-2017/24/C/ST3/00276 (K.J.K.) and UMO-2017/25/B/ST3/02586 (A.P.).
K.J.K. thanks the Polish National Agency for Academic Exchange for funding in the frame of the Bekker programme (PPN/BEK/2020/1/00184).
In addition, K.J.K. and A.P.  are grateful for the funding from the scholarships of the Minister of Science and Higher Education (Poland) for outstanding young scientists (2019 edition, Nos.~821/STYP/14/2019 and 818/STYP/14/2019, respectively).
Access to  computing and storage facilities provided by the Poznan Supercomputing and Networking Center (EAGLE cluster) is greatly appreciated.
\end{acknowledgments}

\bibliography{bibliography}

\end{document}